\documentclass[preprint,review,12pt]{elsarticle}
\usepackage{amssymb}
\usepackage[T1]{fontenc}
\usepackage{graphics}
\usepackage{bm}
\usepackage{subcaption}
\usepackage{lmodern}
\usepackage{amsmath,amsxtra,amssymb,latexsym, amscd,color}
\usepackage[mathscr]{eucal}
\def\pA{proton-nucleus\ }
\def\nA{nucleon-nucleus\ }
\def\pG{$(p,\gamma)$\ }
\def\C3p{$p+^{13}$C\ }
\def\pC23{$p+^{12,13}$C\ }
\def\CpG{$^{12,13}$C$(p,\gamma)$\ }
\journal{Nuclear Physics A}
\begin{document}
\begin{frontmatter}
\title{Folding model approach to the elastic \pC23 scattering at low energies and 
 radiative capture \CpG reactions}
	\author{Nguyen Le Anh}\address{Ho Chi Minh City University of Education\\
			280 An Duong Vuong, District 5, Ho Chi Minh City, Vietnam.}
	\author{Nguyen Hoang Phuc	\footnote{Corresponding author: 
	  {\it hoangphuc@vinatom.gov.vn}}, Dao T. Khoa} 
	\address{Institute of Nuclear Science and Technology, VINATOM\\
			179 Hoang Quoc Viet, Cau Giay, Hanoi, Vietnam.}
	\author{Le Hoang Chien, Nguyen Tri Toan Phuc}
	\address{University of Science, VNU-HCM \\
	227 Nguyen Van Cu, District 5, Ho Chi Minh City, Vietnam.}
\begin{abstract}
The proton radiative capture $^{12,13}$C$(p,{\gamma})$ reactions at astrophysical 
energies, key processes in the CNO cycle, are revisited in the potential model
with the \pA potential for both the scattering and bound states obtained in the 
folding model, using a realistic density dependent nucleon-nucleon interaction. 
For the consistency, this same folding model is also used to calculate the optical 
potential of the elastic \pC23 scattering at energies around the Coulomb barrier. 
The folded \pC23 optical potentials are shown to account well for both the elastic 
\pC23 scattering and astrophysical $S$ factors of the radiative capture \CpG reactions.  
\end{abstract}
\end{frontmatter}
\section{Introduction}
\label{intro}
The energy production in stars more massive than the Sun is predominantly generated 
by the hydrogen burning process in the CNO cycle \cite{Bet39}. In particular,
the $^{12}\text{C}(p,\gamma){}^{13}\text{N}$ and $^{13}\text{C}(p,\gamma){}^{14}\text{N}$ 
reactions are the first radiative capture reactions in the CNO cycle that convert 
hydrogen to helium in the hot stellar medium. These reactions not only play a key 
role in the energy production in massive stars but also control the buildup 
of $^{14}{\rm N}$ nuclei in the three-step process 
$^{12}{\rm C}(p,\gamma)^{13}{\rm N}(\beta^+,\nu)^{13}{\rm C}(p,\gamma)^{14}{\rm N}$, 
which governs the $^{12}{\rm C}/{}^{13}{\rm C}$ ratio. The carbon abundance ratio 
is an important characteristic of the stellar evolution and nucleosynthesis \cite{Cla83}. 
These reactions are also believed to be a source of solar neutrinos \cite{Dav94}. 
The production of $^{13}$C by the 
$^{12}{\rm C}(p,\gamma)^{13}{\rm N}(\beta^+,\nu)^{13}{\rm C}$ reactions also 
provides the seed for the $\alpha$ capture reaction $^{13}{\rm C}(\alpha,n)^{16}{\rm O}$, 
the main neutron source for the production of heavier isotopes in the asymptotic giant 
branch (AGB) stars \cite{Bus01}. As a result, the knowledge of the reaction rates 
of the \CpG reactions is very essential for the network calculation of the stellar 
nucleosynthesis. This work is aimed at suggesting a reliable mean-field potential
approach to study these two important reactions.  

In general, the compound nucleus formation and direct capture are the two competing 
mechanisms of a proton-induced reaction at astrophysical energies. However, for 
the \CpG reactions at the sub-barrier energies, the capture process is much more 
dominant because of a small number of excited states of the compound nucleus. Thus, 
the direct radiative capture \CpG process is mainly associated with the electromagnetic 
transition from an initial scattering state to the ground state (g.s.) of the daughter 
nucleus, and a microscopic description of this process should be based on the solutions 
of the nuclear many-body problem for the involved states of the \pA system \cite{Desc20}. 
However, this approach is rather complicated because of the full antisymmetrization 
of the total nuclear wave function. Two other widely used approaches 
to deal with this problem are the phenomenological $R$-matrix method 
\cite{Lane58,Desc10} and potential model \cite{Ber03,Hua10,NACRE2}. 

An advantage of the phenomenological $R$-matrix method is its ability to reproduce 
the experimental data with high precision. This method depends, however, on the 
available data and cannot be used to predict the nuclear reactions involving the 
short-lived unstable nuclei. The $R$-matrix method has been used to describe \CpG 
reactions using the asymptotic normalization coefficient of the $p+^{12}$C and 
$p+^{13}$C systems inferred from the DWBA analysis of the nucleon transfer reactions 
\cite{Bur08,Muk03,Art08} or treated as a free parameter \cite{Cha15}. 

The potential model has been proven to be a convenient tool to study the radiative 
capture reactions, especially, those involving light nuclei. The potential model was used, 
in particular, for the \CpG reaction using phenomenological \pA potential (see Review 
by Huang {\it et al.} \cite{Hua10} as well as the NACRE II evaluation \cite{NACRE2}). 
The potential model was also combined with the phenomenological $R$-matrix method 
to estimate the radiative decay width of the $^{13}$N$^*$ resonance \cite{Muk17}. 
Instead of using phenomenological potentials, the microscopic \pA potential can be  
calculated in the folding model \cite{Kho02} using a realistic effective nucleon-nucleon 
(NN) interaction and the appropriate single-particle wave functions of target nucleons. 
The use of the folded \pA potential not only reduces the number of free parameters, 
but it also shows the important nuclear mean-field aspect of the radiative capture 
\pG reactions and validates eventually the use of the folded \nA potential to predict 
reaction rates of the nucleon-induced reactions with unstable nuclei. 

The folding model of the \pA potential was used earlier with the density independent M3Y 
interaction \cite{Ber77} to study the $(p,\gamma)$ reactions occurring in the CNO cycle, 
for example, the recent determination of the direct capture component of the 
$^{13}$C$(p,{\gamma})$ reaction  using the folded $p+^{13}$C potential \cite{Cha19}. 
In the present work, the local \pA optical potential (OP) is calculated using the recent 
(mean-field based) version \cite{Loan15} of the folding model \cite{Kho02} and used 
consistently in both the optical model (OM) analysis of the elastic \pC23 scattering at 
low energies and potential model study of the \CpG reactions. The density dependent 
CDM3Y3 interaction used in the present folding model calculation is the original 
M3Y-Paris interaction \cite{M3Y-Paris} supplemented by a realistic density dependence 
\cite{Loan15,Kho97}. This version of the in-medium NN interaction has been proven 
to give an accurate description of the saturation properties of cold nuclear matter 
(NM) within the Hartree-Fock (HF) formalism \cite{Kho93,Than09,Loan11} as well as 
the \nA and nucleus-nucleus OP's for the OM analysis of elastic nucleon- and heavy-ion 
scattering \cite{Kho02,Loan15,Kho97,Kho07,Kho09}. In the recent HF study of the 
single-particle  potential in NM \cite{Loan15}, the density dependence of the 
CDM3Y3 interaction has been further modified to include the rearrangement term (RT) 
of the single-particle potential in NM limit given by the Hugenholtz-van Hove 
theorem. The inclusion of the RT into the folding calculation of the nucleon OP 
was shown to be essential for a good OM description of elastic nucleon scattering 
at low energies \cite{Loan15,Loan20}, and it is expected to provide also a reliable 
description of the \pG reactions of light nuclei. 

\section{Potential model of the radiative capture reaction}
\label{sec2}
For the nuclear reaction induced by proton at an energy $E$ well below the Coulomb 
barrier, the reaction cross section $\sigma(E)$ decreases too rapidly with 
the decreasing energy, and it is convenient to investigate the astrophysical 
$S$ factor of the proton radiative capture reaction determined as 
\begin{equation} 
 S(E) = E\exp(2\pi\eta)\sigma(E), \label{eq1}
\end{equation}
where the Sommerfeld parameter $\eta=Z e^2/(\hbar v)$, $Z$ is the atomic number
of the target, and $v$ is the proton (relative) velocity. In the potential 
model of the $(p,\gamma)$ reaction, the bound or scattering state is obtained 
from the solution of the radial Schr\"odinger equation 
\begin{equation} 
\left\{\dfrac{d^2}{dr^2} -\dfrac{\ell(\ell+1)}{r^2} +\dfrac{2m}{\hbar^2} 
\left[E - V(r)\right]\right\} \psi(r) = 0, \label{eq2}
\end{equation} 
where $\ell$ is the orbital angular momentum, $V(r)$ is the total \pA potential 
that includes the nuclear $V_{\rm N}$, spin-orbit $V_{\rm SO}$ and Coulomb $V_{\rm C}$ 
potentials; $m$ is the reduced mass, and $E$ is the proton energy in the center of mass 
(c.m.) frame. We further denote the solution $\psi(r)$ for the scattering state ($E>0$) 
as $\chi(r)$, and that for the bound state ($E<0$) as $\phi(r)$.

With the energy of the (initial) scattering state expressed in terms of the wave 
number $k$ as $E ={\hbar^2 k^2}/{(2m)}$, the asymptotic scattering wave function 
(at large distances where the nuclear potential $V\to 0$) is 
\begin{equation}
\chi_{\ell_i j_i}(k,r\to \infty) \to [F_{\ell_i}(kr) \cos\delta_{\ell_i j_i} 
+G_{\ell_i}(kr) \sin\delta_{\ell_i j_i}]\exp[i(\delta^c_{\ell_i}+
\delta_{\ell_i j_i})], \label{eq3}
\end{equation}
where $G$ is the regular- and $F$ is the irregular Coulomb functions \cite{Abra65}. 
$\delta^c_{\ell_i}$ and $\delta_{l_i j_i}$ are the Coulomb and nuclear phase shifts, 
respectively. 

The wave function of the (final) bound state is negligible at large distances, 
and its norm is determined as
\begin{equation}
\int_0^\infty |\phi_{n_f \ell_f j_f}(r)|^2 dr =1, \label{eq4}
\end{equation}
where $n_f$ is the node number of $\phi(r)$. In the one-channel potential 
model, both the scattering and bound states in the \pG reaction are generated by 
the appropriately chosen \pA potential. A convenient choice for this purpose is 
the phenomenological Woods-Saxon (WS) potential, and sometimes one has to use two 
different WS potentials for the scattering- and bound states to obtain a good description 
of the \pG cross section \cite{Hua10}. In the present work, we use the \pA potential 
predicted consistently by the same (mean-field based) folding model \cite{Loan15,Loan20} 
to determine both $\chi$ and $\phi$ from the solutions of the Schr\"odinger 
equation (\ref{eq2}). 

The electric transition matrix element between $\chi$ and $\phi$ is readily determined 
by the Fermi golden rule, using the long wave-length approximation. The total 
cross section of the \pG reaction is obtained by summing the partial cross
sections over all initial states $|J_i M_i\rangle$, final states $|J_f M_f\rangle$, 
and electric multipoles $\lambda$. 
\begin{equation}
\sigma(E)=\sum_{\lambda J_i J_f }\sigma_{\lambda,J_i\to J_f}(E). \label{eq5}
\end{equation}
\begin{equation}\label{eq6}
\sigma_{\lambda,J_i\to J_f}(E)=\frac{4\pi(\lambda+1)}{\lambda\left[(2\lambda+1)!!
\right]^2(2S+1)}\frac{m c^2}{(\hbar c)^2}\frac{k_\gamma^{2\lambda+1}}{k}
S_{\text{F}}\frac{|\langle J_f||\hat{O}_{\lambda}||J_i \rangle |^2}{2J_i+1},
 \end{equation}
where $|J_i \rangle $ and $|J_f\rangle$ are the total wave functions of the 
initial and final states, respectively; $S$ is spin of the target (or the core
of daughter nucleus), and $S_{\text{F}}$ is the often dubbed as the spectroscopic 
factor of the final state. The wave numbers $k$ and $k_\gamma$ are determined by 
the proton energy $E$ and that of the emitted photon $E_\gamma$, respectively. 
The multipole component of the electric transition operator is determined as
\begin{equation}\label{eq7}
\hat{O}_{\lambda\mu}=\frac{e[A^\lambda+(-1)^\lambda Z]}{(A+1)^\lambda}
 r^\lambda Y_{\lambda\mu}(\hat{\bm r})=C_\lambda r^\lambda Y_{\lambda\mu}(\hat{\bm r}),
\end{equation}
where $A$ is the mass number of the target. In the present work, the total wave 
function of the initial (final) state in Eq.~\eqref{eq6} is determined from the 
coupling of the incident (valence) proton to the target (core) nucleus as
\begin{equation}\label{eq8}
 |J_i\rangle = \left[\chi_{\ell_ij_i}\otimes\Psi_S\right]_{J_i}\ \mbox{and}\  
 \ |J_f\rangle = \left[\phi_{n_f\ell_fj_f}\otimes\Psi_S\right]_{J_f}.
\end{equation}
After integrating out the angular dependences of the wave functions, the total 
cross section of the \pG reaction is obtained as 
\begin{align}\label{eq9}
\sigma_{\lambda,J_i \to J_f}(E)=\frac{4\pi(\lambda+1)(2\lambda+1)}{\lambda
\left[(2\lambda+1)!!\right]^2}\frac{mc^2}{(\hbar c)^2}\frac{k_\gamma^{2\lambda+1}}
{k^3}C_\lambda^2\frac{(2J_i+1)(2J_f+1)}{(2S+1)}S_{\text{F}} \nonumber\\
 \times\sum_{\ell_i,j_i}\Big\vert i^{\ell_i}(-1)^{j_f+\ell_f}\hat{j_i}\hat{j_f}\hat{\ell_i}
 \left\{\begin{matrix}
   j_i&J_i& S  \\
   J_f& j_f&\lambda
  \end{matrix} \right\}
	\left\{\begin{matrix}
   \ell_i&j_i& \frac{1}{2}  \\
   j_f& \ell_f&\lambda
 \end{matrix}\right\}\langle \ell_i0,\lambda0\vert \ell_f0 \rangle I(k)
 \Big\vert^2 ,
 \end{align}
where $\hat{j}=\sqrt{2j+1}$, and the radial overlap of the scattering- and 
bound proton states is 
\begin{equation} \label{eq10}
I(k)=\int_0^\infty\phi_{n_f\ell_fj_f}(r)\chi_{\ell_i j_i}(k,r)r^\lambda dr. 
\end{equation}
Thus, the most vital nuclear physics input of the scattering- and bound proton 
states of the \pG reaction is embedded in the overlap integral \eqref{eq10} 
and spectroscopic factor $S_{\text{F}}$ of the final nuclear state. As usually 
adopted in the potential model analysis of the \pG reactions \cite{Hua10}, the 
$S_{\text{F}}$ value is obtained from the best fit of the calculated \pG cross 
section to the measured data. The cross sections of the \CpG reactions under study 
are determined mainly by the electric dipole ($\lambda=1$) transition from the 
\pC23 scattering states to the ground states of ${}^{13,14}{\rm N}$ nuclei that 
are approximately treated in terms of the valence $1{\rm p}_{\frac{1}{2}}$ proton 
coupled to the inert $^{12,13}{\rm C}$ cores, with the spins of $^{13}$N and 
$^{14}$N are $J^\pi_f=\frac{1}{2}^{-}$ and $1^{+}$, respectively. 

In general, both the resonance and nonresonant \pG processes contribute to 
the radiative capture cross section (\ref{eq5}). While the resonance scattering wave 
function is determined by the \pA potential accurately fine tuned to reproduce the 
peak of resonance in the solution of Eq.~(\ref{eq2}), the nonresonant contribution 
to the \pG cross section is spread over a wide range of energies $E$ and correlates 
closely with the asymptotics of the final (bound) state wave function expressed 
explicitly via Whittaker function \cite{Muk97} as
\begin{equation}\nonumber
\phi_{n_f\ell_fj_f}(r)=b_{n_f\ell_fj_f}W_{-\eta,\ell_f+1/2}(2kr),\ {\rm with}\ r>R_N.
\end{equation}
Here $b_{n_f\ell_fj_f}$ is the amplitude of the asymptotic tail of the bound state 
wave function which is dubbed as the single-particle Asymptotic Normalization 
Coefficient (ANC), $k$ is the wave number of the bound state and radius $R_N$ 
is chosen large enough that the nuclear potential at $r>R_N$ is negligible. 
It is obvious that the nonresonant \pG cross section is proportional to 
$S_{\text{F}} b^2_{n_f\ell_fj_f}$. In the potential model study of the \pG 
reaction, different choices of the proton binding potential for the final state wave 
function might well lead to different $b_{n_f\ell_fj_f}$ values. Therefore, it 
is more appropriate to use the final state ANC determined \cite{Muk97} as 
\begin{equation}
  A_{\text{F}} = S_{\text{F}}^{1/2}b_{n_f\ell_fj_f}. \label{eq10a}
\end{equation}
In the ``valence nucleon + core'' model used in the present work for the daughter 
nucleus, the spectroscopic factor of the final bound state is that of the valence 
nucleon, i.e., $S_{\text{F}}\approx S_{n_f\ell_fj_f}$ which can be taken from the
prediction of the shell model or deduced from the analysis of the nucleon transfer 
and/or breakup reactions. The final state ANC (\ref{eq10a}) is thus more or less
independent on the binding potential of the valence nucleon. 

We finally note that if we choose to couple the proton spin to that of the target 
or core nucleus to a total nuclear spin $\bm{I}_{i(f)}=\bm{\frac{1}{2}}+\bm{S}_{i(f)}$, 
then instead of Eq.~(\ref{eq9}) we end up with the same expression for the total \pG cross 
section as that given in NACRE II compilation \cite{NACRE2} which is often used in potential 
model studies. However, in this way the total angular momentum $j_{i(f)}$ of the incident 
(bound) proton is \emph{not} explicitly determined. As a result, the spectroscopic 
information on the single-particle configuration $n_f\ell_fj_f$ of the valence proton 
is lost.     

\section{Nuclear mean-field potential and the folding model}
\label{sec3}
The total \pA potential $V(r)$ in Eq.~\eqref{eq2} contains the nuclear mean-field
part (central part), the spin-orbit (SO) and Coulomb (C) potentials as
\begin{equation}\label{eq11}
V(r)=V_{\rm N}(r)+V_{\rm SO}(r)(\bm{\ell}\cdot\bm{\sigma})+V_{\rm C}(r).
\end{equation}
The Coulomb potential of a uniformly charged sphere is used in the OM
calculation of elastic proton scattering and potential model calculation
of the \pG reaction
\begin{eqnarray}
V_{\rm C}(r)=\left\{\begin{array}{lcr}
\displaystyle\frac{Ze^2}{r},& & r > R_{\rm C}\\
\displaystyle\frac{Ze^2}{2R_{\rm C}}\left(3-\frac{r^2}{R_{\rm C}^2}\right),& & 
 r\leqslant R_{\rm C}, \label{eq12}
\end{array}\right. 
\end{eqnarray}
where the Coulomb radius $R_{\rm C}= r_{\rm C}A^{1/3} = 1.25~A^{1/3}$ (fm).

In the phenomenological potential model studies of the \pG reactions (see, e.g., 
Ref.~\cite{Hua10}), the WS form is mostly used for the central and spin-orbit 
potentials in Eq.~\eqref{eq10}
\begin{align} 
V_{\rm N}(r) &= -V_{\rm N}f_{\rm N}(r), \label{eq13}\\
V_{\rm{SO}}(r) &= V_{\rm SO} \left(\dfrac{\hbar}{m_\pi c}\right)^2 
\dfrac{1}{r} \dfrac{d}{dr} f_{\rm SO}(r),\label{eq14} \\
{\rm where}\ f_x(r) &=\left[1+\exp\left(\dfrac{r-R_x}{a_x}\right)\right]^{-1}, 
\ x={\rm N},{\rm SO}. \label{eq15}
\end{align}
Here $R_x$ and $a_x$ are the radius and diffuseness parameters of the WS potential, 
respectively. The squared pion Compton wavelength $[\hbar/(m_\pi c)]^2\approx 2$ fm$^2$ 
is frequently used in Eq.~\eqref{eq14} so that $V_{\rm SO}$ can be given in MeV.

\subsection*{Folding model of the \pA potential}
The folding model of nucleon OP predicts the first-order term of the microscopic 
nucleon OP within the Feshbach's formalism of nuclear reactions \cite{Fe92}. 
The folded nucleon OP has been proven to be successful in the OM description 
of elastic \nA scattering at low and medium energies \cite{Kho02,Loan20}.
In the present work, we extend the use of the folding model to the low energies 
of nuclear astrophysics. In general, the proton OP is evaluated \cite{Kho02,Loan15,Loan20} 
as the following HF-type potential
\begin{equation} 
  V_{\rm fold}=\sum_{j\in A}[\langle pj|v_{\rm D}|pj\rangle 
	+\langle pj|v_{\rm EX}|jp\rangle], \label{eq16}
\end{equation}
where $v_{\rm D(EX)}$ are the direct and exchange parts of the effective NN 
interaction between the incident proton $p$ and bound nucleon $j$ of the target 
$A$. For the NN interaction, we have used the CDM3Y3 density dependent version 
\cite{Loan15,Kho97} of the M3Y-Paris interaction \cite{M3Y-Paris}, with its 
density dependent parameters originally determined \cite{Kho97} to reproduce 
the saturation properties of symmetric NM and recently modified to include
the rearrangement effect of the single-nucleon potential in the folding
calculation \cite{Loan15,Loan20}. 

The direct term of the proton OP can be expressed in terms of the isoscalar 
(IS) and isovector (IV) parts \cite{Loan20} as   
\begin{align} 
 V_{\rm D}(r)&=V^{\rm D}_{\rm IS}(r)+V^{\rm D}_{\rm IV}(R), \nonumber\\ 
 V^{\rm D}_{\rm IS(IV)}(r)&=\int\big[\rho_p(\bm{r}')\pm\rho_n(\bm{r}')\big]
 v^{\rm D}_{00(01)}(\rho,s)d^3r',\ s=|\bm{r}-\bm{r}'|, \label{eq17}  
\end{align}
where $\rho_\tau(\bm r)$, with $\tau=p$ and $n$, are the proton and neutron g.s. 
densities, respectively. The ($+$) sign pertains to the IS part and ($-$) sign 
to the IV part of the folded potential. The antisymmetrization of the \pA system 
is done in the HF manner \cite{Loan20} which leads to a \emph{nonlocal} exchange 
term of the folded OP (\ref{eq16}). Using the WKB approximation for the shift 
of the scattering wave by the spatial exchange of the incident proton and nucleon 
bound in the target \cite{Bri77}, the exchange term in Eq.~(\ref{eq16}) becomes 
localized as   
\begin{align}
 V_{\rm EX}(E,r)&=V^{\rm EX}_{\rm IS}(E,r)+V^{\rm EX}_{\rm IV}(E,r), \nonumber\\ 
 V^{\rm EX}_{\rm IS(IV)}(E,r)&=\int\big[\rho_p(\bm{r},\bm{r}')\pm\rho_n(\bm{r},\bm{r}')\big] 
  j_0\big(k(E,r)s\big)v^{\rm EX}_{00(01)}(\rho,s)d^3r'.	\label{eq18}
\end{align}
Here $v^{\rm D(EX)}_{00}(\rho,s)$ and $v^{\rm D(EX)}_{01}(\rho,s)$ are the 
(density dependent) IS and IV components of the direct and exchange parts 
of the central CDM3Y3 interaction \cite{Loan15}. 
As a result, the proton OP depends explicitly on energy via the proton relative 
momentum which is determined self-consistently from the folded \pA potential 
$V_{\rm fold}(E,r)=V_{\rm D}(r)+V_{\rm EX}(E,r)$ as
\begin{equation}
 k^2(E,r)=\frac{2m}{\hbar^2}[E-V_{\rm fold}(E,r)-V_{\rm C}(r)]. \label{eq19}
\end{equation}  
The readers are referred to Refs.~\cite{Kho02,Loan20} for more details of the 
folding model calculation using the \emph{finite-range} exchange $v^{\rm EX}_{00(01)}$ 
interaction. Applying a local approximation \cite{Campi78} to the nonlocal g.s. 
density matrices in the exchange integral (\ref{eq18}), the local g.s. density 
given by any nuclear structure model can be used in the folding model calculation 
of the \pA OP. 

Given a high sensitivity of both the proton binding energy and resonance peak 
to the strength of the nuclear potential, a renormalization of the folded 
\pA potential is introduced and adjusted to reproduce the resonance peak 
of the $(p,\gamma)$ reaction and binding energy of the valence proton in 
the daughter nucleus   
\begin{equation}
V_{\rm N}(E,r) = N_i V_{\rm fold}(E,r), \label{eq20}
\end{equation}
where $i=s$ and $b$ for the scattering (resonance) and bound proton states, 
respectively. In general, the use of the \pA folded potential for either the 
proton OP or binding potential to study the \pG reaction is appropriate when the 
best-fit $N_i$ factor is close to unity. For the resonance and nonresonant 
scattering states, the \pA potential is calculated over the whole energy range 
under study, with the step $\Delta E=5$ keV. For the proton bound state, the 
experimental binding energy of the valence proton in the daughter nucleus is used
as the input ($E=-E_b$) in the folding calculation (\ref{eq17})-(\ref{eq18}) 
of the \pA potential.     

Besides an effective (density dependent) NN interaction, a proper choice of neutron 
and proton densities $\rho_\tau(r)$ of the target is an important input for the 
folding model calculation. In the present work, the g.s. densities of $^{12,13}$C 
nuclei were obtained in the independent particle model (IPM) \cite{Sat79} as 
\begin{equation} 
\rho_\tau(r)=\dfrac{1}{4\pi}\sum_{n\ell j}S_{n\ell j\tau}\left(\dfrac{A}{A-1}\right)^3 
 \phi^2_{n\ell j\tau}\left(\dfrac{A}{A-1}r\right),\ \tau=n\ {\rm and}\ p, \label{eq21}
\end{equation}
where the factor $A/(A-1)$ accounts for the recoil effect. Each single-nucleon wave 
function $\phi_{n\ell j\tau}$ is determined from Eq.~\eqref{eq2} using a separate WS 
potential (\ref{eq13})-(\ref{eq15}). The WS depth was adjusted in each case to give 
the required binding energy using a fixed WS diffuseness $a=0.65$ fm, and the WS 
radius fine tuned for the proton g.s. density to have the root-mean-squared radius 
$R_p$ close to that deduced from the measured charge radius of $^{12}$C or $^{13}$C. 
The nucleon binding energies and spectroscopic factors $S_{n\ell j\tau}$ for target
nucleons were taken from the shell model results (see Ref.~\cite{Sat79} for more 
detail). Within the IPM, $^{13}$C is treated as the $^{12}$C core coupled 
to the valence $1{\rm p}_{1/2}$ neutron moving in the same WS well \cite{Sat79}. 

\section{Elastic \pC23 scattering at low energies}
\label{sec4}
\begin{figure}\vspace*{-1cm}
\centerline{\includegraphics[width=\textwidth]{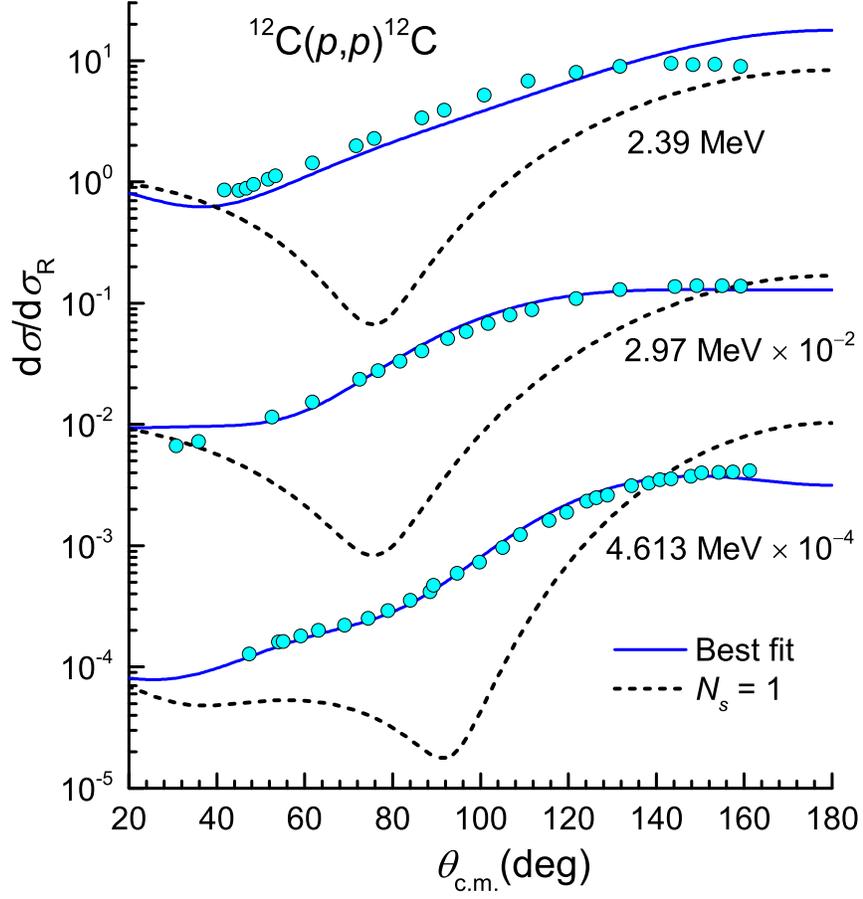}}\vspace*{-0.5cm}
\caption{OM description of the elastic $p+^{12}$C scattering data measured 
at proton incident energies of $2.39, 2.97$, and $4.61$ MeV \cite{p12Cdata,p12Cdata1} 
given by the folded proton OP. The best OM fit (solid lines) is given by the folded 
potential renormalized by $N_s=1.32$ at $E_p=2.39$ MeV, and $N_s=1.33$ at $E_p=2.97$ 
and 4.61 MeV.} \label{f1} 
\end{figure}

\begin{figure}\vspace*{-1cm}
\centerline{\includegraphics[width=\textwidth]{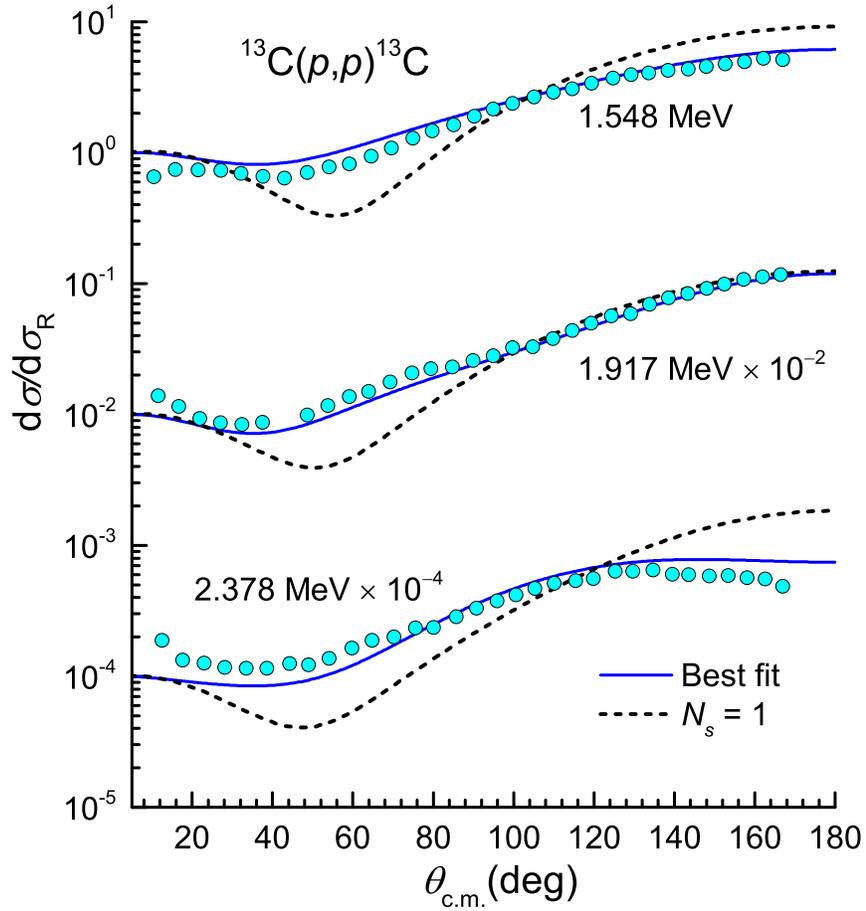}}\vspace*{-0.5cm}
\caption{The same as Fig.~\ref{f1} but for the elastic $p+^{13}$C scattering data 
measured at proton incident energies of 1.55, 1.92, and 2.38 MeV \cite{p13Cdata}. 
The best fit (solid lines) is given by the folded potential renormalized by 
$N_s=1.23$, 1.21, and 1.25 at $E_p=1.55$, 1.92 MeV, and 2.38 MeV, respectively.} 
 \label{f2} 
\end{figure}

To validate the folded \pC23 potentials for the potential model study of the
$^{12,13}{\rm C}(p,\gamma)$ reactions at sub-barrier energies, we found it necessary 
to test the folded proton OP first in the OM analysis of elastic \pC23 scattering 
at low energies. Namely, the elastic $p+^{12}$C data measured at proton energies 
of 2.39, 2.97, and 4.61 MeV \cite{p12Cdata,p12Cdata1}, and the elastic $p+^{13}$C data 
at 1.548, 1.917, and 2.378 MeV \cite{p13Cdata} were compared with the OM results 
given by the folded \pA OP (see Figs.~\ref{f1} and \ref{f2}). All OM calculations 
were done using the code ECIS97 written by Raynal \cite{Ray72}. 
 
It is noteworthy that these energies are below the thresholds of the inelastic 
$(p,p')$ scattering and $(p,d)$ reaction, and the imaginary OP caused by the 
(on-shell) coupling of the elastic scattering channel to these nonelastic 
channels can be neglected \cite{Bran97}. Thus, the real folded \pC23 potential 
is used as the total OP for these systems in our OM calculation, 
$V_{\rm N}=N_sV_{\rm fold}(r)$ with $N_s$ adjusted to the best OM fit to  
elastic data. The WS spin-orbit potential (\ref{eq14}) is constructed using 
$V_{\rm SO}=5$ MeV, $a_{\rm SO}=0.65$ fm, and $R_{\rm SO}\approx 2.9$ fm for 
the OM calculation in both cases. One can see in Figs.~\ref{f1} and \ref{f2}
that a good OM description of the considered elastic data has been obtained 
with the folded proton OP renormalized by $N_s\approx 1.32$--1.33 for the 
$p+^{12}$C system and $N_s\approx 1.20$--1.25 for the $p+^{13}$C system. 
With a rather weak Coulomb interaction, the contribution of the nuclear potential 
to elastic \pC23 scattering is still very significant at these energies, 
and the elastic cross section at medium and large angles is by a factor of 10 
or more stronger than the corresponding Rurtheford cross section (see Figs.~\ref{f1} 
and \ref{f2}). 

As a result, the measured elastic data are quite sensitive to the strength 
of $V_{\rm N}$. The unrenormalized folded OP with $N_s=1$ fails to reproduce the 
data at medium and large angles (see dash lines in Figs.~\ref{f1} and \ref{f2}). 
The agreement with the data cannot be improved when we used a complex OP consisting 
of the unrenormalized real folded potential and the WS imaginary potential with  
parameters being adjusted to fit the OM results to the measured data. Within the 
frame of the microscopic OP \cite{Fe92}, a renormalization factor $N_s$ larger 
than unity by 20--30 \% is caused by the higher-order (beyond mean-field) 
contributions to the folded \pC23 real OP, like the off-shell coupling of the
elastic channel to the closed nonelastic channels \cite{Bran97} and/or the 
effect of the nonlocality. This should be the subject of the further folding model 
study of the elastic \pC23 scattering at barrier energies.  

\section{Radiative capture $^{12,13}\text{C}(p,\gamma)$ reactions}
\subsection{Results and discussion for the $^{12}{\rm C}(p,\gamma)^{13}{\rm N}$ 
 reaction} \label{sec5.1}
The direct radiative capture $^{12}{\rm C}(p,\gamma)$ to the g.s. of $^{13}$N 
is proven to be dominated by the single-proton $E1$ transition from the s$_{1/2}$ 
scattering state to the 1p$_{1/2}$ bound state of $^{13}$N. In the present work, the 
folded $p+^{12}$C potential (\ref{eq20}) has been used in Eq.~(\ref{eq2}) to generate 
the bound-state wave function $\phi_{1{\rm p}_{1/2}}$. Using the same spin-orbit potential 
as given above in Sect.~\ref{sec4} for both the scattering and bound states, the folded 
$p+^{12}$C potential calculated at $E=-E_b$ needs to be renormalized by $N_b\approx 1.005$ 
for the radial Schr\"odinger equation (\ref{eq2}) to reproduce the binding energy 
$E_b\approx 1.943$ MeV of the valence 1p$_{1/2}$ proton of $^{13}$N. Such a $N_b$ factor 
validates the use of the current version of the folding model to generate the wave
function of the nucleon bound states in the potential model study of the 
\pG reaction.         

\begin{figure}[th]\vspace*{-1.0cm}
\centerline{\includegraphics[width=\textwidth]{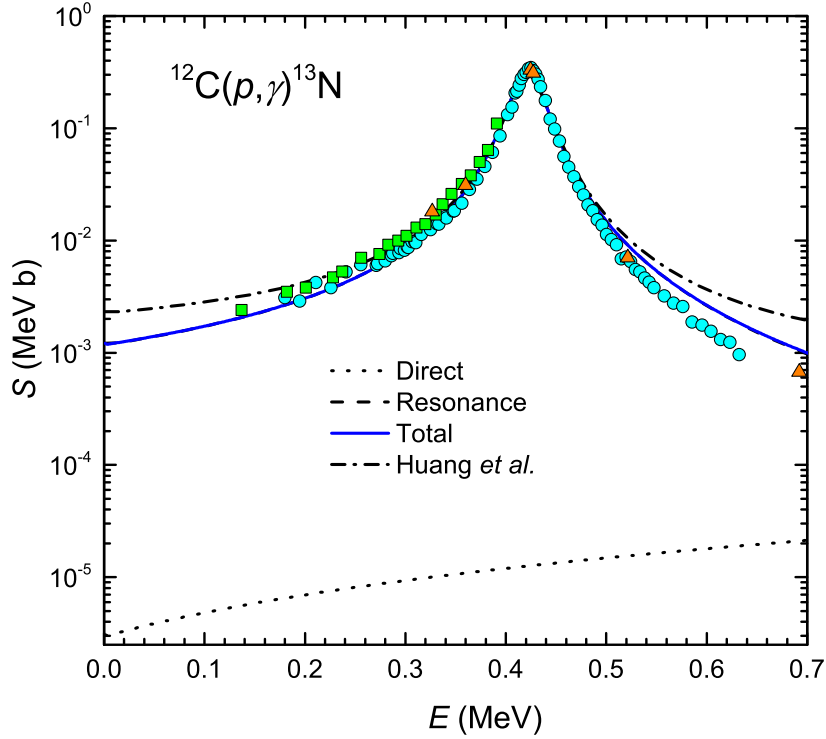}}\vspace*{-0.5cm}
\caption{Astrophysical $S$ factor (\ref{eq1}) of the $^{12}{\rm C}(p,\gamma)^{13}{\rm N}$ 
reaction given by the folded $p+^{12}$C potential using the spectroscopic factor 
$S_{\rm F}\approx 0.40$ and 0.61 for the resonance and direct components, 
respectively. The experimental data shown as triangles, circles, and squares 
were taken from Refs.~\cite{Bur08}, \cite{Vogl63}, and \cite{Rol74}, 
respectively.} \label{f3} 
\end{figure}
The $1/2^+$ resonance peak of the $^{12}$C$(p,\gamma)$ reaction has been observed 
in different experiments at $E\approx 0.422$ MeV above the proton threshold 
of $^{13}$N \cite{Bur08,Vogl63}. Because the resonance energy (the location of the 
resonance peak) is determined directly by the OP for the proton s$_{1/2}$ scattering 
wave function, we adjusted the normalization factor $N_s$ to reproduce the 
resonance peak at $E_R \approx 0.422$ MeV and obtained the best fit $N_s\approx 1.317$, 
with the statistic $\chi^2\approx 0.021$ for one degree of freedom. In agreement 
with the OM results discussed above, the good potential model description of the 
resonance scattering state requires the strength of the folded $p+^{12}$C potential 
to be increased by about 30\%. The spectroscopic factor $S_{\text{F}}\approx 0.40$ 
for the resonance cross 
section was given by the best fit of the calculated astrophysical $S$ factor 
(\ref{eq1}) to the measured data \cite{Bur08,Vogl63,Rol74}. The energy 
dependent astrophysical factors $S(E)$ given by the folded $p+^{12}$C potential 
and phenomenological WS potential \cite{Hua10} using $S_{\rm F}\approx 0.40$ and 
0.35, respectively, agree well with the measured data \cite{Bur08,Vogl63,Rol74} 
(see Fig.~\ref{f3}). Usually one might tend to assign the best-fit $S_{\text{F}}$ 
value to the spectroscopic factor of the proton 1p$_{1/2}$ state in the g.s. of $^{13}$N, 
but it is just a factor scaled to reproduce the peak of the $1/2^+$ resonance in 
the cross section $\sigma(E)$ or astrophysical factor $S(E)$ of the $^{12}$C$(p,\gamma)$ 
reaction. Although $S_{\rm F}$ is correlated with the strength of the bound proton 
state through the overlap integral (\ref{eq9}), it should not be compared with 
the spectroscopic factor extracted from the one-proton knockout (pickup) reaction 
\cite{Tsa05} or predicted by the \emph{ab initio} shell model calculations \cite{Tim13} 
which is directly associated with the occupancy probability of the valence nucleon. 
For the proton 1p$_{1/2}$ bound state of $^{13}$N the spectroscopic factor 
$S_{{\rm 1p}_{1/2}}\approx 0.59\sim 0.63$ given by the \emph{ab initio} calculation 
\cite{Tim13} is about 50\% larger than $S_{\rm F}\approx 0.40$ obtained from the 
present potential model study of the $^{12}{\rm C}(p,\gamma)$ reaction.    
\begin{figure}[th]\vspace*{-1cm}
\centerline{\includegraphics[width=\textwidth]{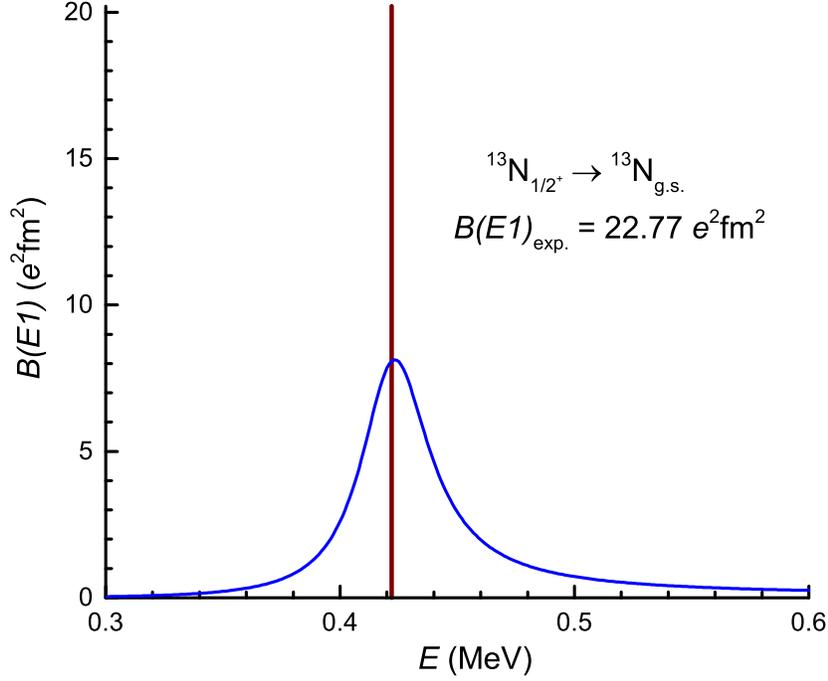}}\vspace*{-0.5cm}
\caption{Dipole transition strength given by the potential model calculation 
of the $^{12}{\rm C}(p,\gamma)^{13}{\rm N}$ reaction, using the folded 
$p+^{12}$C potential and spectroscopic factor $S_{\rm F}\approx 0.40$
(solid line). At the resonance peak, the calculated reduced transition probability
$B(E1;1/2^+ \to 1/2^-)\approx 20.31\ e^2$~fm$^2$ (vertical bar), in comparison
with the measured value of $22.77\pm 0.25\ e^2$~fm$^2$ \cite{Endt93}.} \label{f4} 
\end{figure}

Technically, the best-fit $S_{\rm F}$ factor obtained in our study of the \pG reaction
is quite helpful in illustrating the spreading of the $E1$ transition strength over 
the energy range covered by the $1/2^+$ resonance as shown in Fig.~\ref{f4}. With 
the reduced $E\lambda$ transition probability given by
\begin{equation}\label{eq22}
 B(E\lambda)=\frac{|\langle J_f ||\hat{O}_{\lambda}||J_i\rangle |^2}{2J_i+1},
\end{equation}
the $E1$ transition strength of the $\gamma$ decay from the $1/2^+$ resonance 
peak to the g.s. of $^{13}$N can be evaluated from the Eqs.~(\ref{eq6}) and 
(\ref{eq9}) using the proton scattering wave function $\chi$ determined at the 
resonance energy and the wave function $\phi$ of the bound (valence) proton, 
with the $^{12}$C core treated as spectator. We have obtained from the calculated
\pG cross section (\ref{eq9}) the reduced transition probability 
$B(E1)=|\langle 1/2^-||\hat{O}_{E1}||1/2^+ \rangle|^2/2\approx 20.31\ e^2$~fm$^2$ 
(vertical bar in Fig.~\ref{f4}) that agrees with the experimental value 
$B(E1)_{\rm exp}\approx 22.77\pm 0.25\ e^2$~fm$^2$ deduced from the $\gamma$ 
transition strength of the $1/2^+$ excited state of $^{13}$N \cite{Endt93}. 
The distribution of the $E1$ strength in the energy range around the $1/2^+$ peak 
obtained from the calculated \pG cross section (\ref{eq9}) is shown in Fig.~\ref{f4}, 
and one can see that the best-fit $S_{\rm F}$ factor helps to smoothly distribute 
the single-proton $E1$ transition strength approximately over twice the width 
of the $1/2^+$ state, $\Gamma_{\rm c.m.}\approx 32$ keV \cite{A-Love91}, as expected. 

In addition to the $E1$ transition from the $1/2^+$ resonance, the nonresonant 
$E1$ transition from the proton d$_{3/2}$ nonresonant scattering states to the 
proton 1p$_{1/2}$ bound state of $^{13}$N is also possible. Such a nonresonant 
contribution is often discussed in literature as the contribution by the direct capture 
\cite{NACRE2}, and the total astrophysical $S$ factor (\ref{eq1}) should be the
sum of both the resonance and nonresonant astrophysical factors. In general, 
the spectroscopic factors $S_{\text{F}}$ of the resonance and nonresonant 
contributions are not necessarily to be the same, and in the ``valence nucleon + core''
model the $S_{\text{F}}$ value for the direct capture cross section can be assumed 
to be that predicted by the shell model or deduced from the nucleon transfer reaction 
\cite{Ili04}, as discussed in Sect.~\ref{sec2}. Thus, the proton spectroscopic 
factor $S_{\text{F}}\approx 0.61$ predicted by the shell model calculation \cite{Coh67}
has been used to obtain the direct capture d$_{3/2}\to {\rm p}_{1/2}$ cross section. 
The $p+^{12}$C folded potential renormalized by the same factor $N_s\approx 1.317$ 
as that obtained for the s$_{1/2}$ resonance scattering state is used to generate 
d$_{3/2}$ nonresonant scattering state. The result obtained for the direct 
capture cross section is shown in terms of the astrophysical $S$ factor as dotted line in 
Fig.~\ref{f3}, one can see that the direct capture does not contribute significantly 
to the total $S$ factor. The final state ANC given by our model calculation is 
$A_{\rm F}^2\approx 2.061$~fm$^{-1}$, which agrees reasonably with those given by 
the phenomenological $R$-matrix method calculation \cite{Bur08} or deduced from 
the proton transfer reaction $^{12}$C($^3$He,$d)^{13}$N \cite{Yar97}, 
$A_{\rm F}^2\approx 2.045$ fm$^{-1}$. Our result is, however, about 50\% lower than 
the value of $A_{\rm F}^2\approx 4.203$ fm$^{-1}$ obtained in the potential model 
analysis by Huang {\it et al.} using the phenomenological \pA potential \cite{Hua10}.  
      
\subsection{Results and discussion for $^{13}{\rm C}(p,\gamma)^{14}{\rm N}$ 
 reaction}\label{sec5.2}
The $^{13}{\rm C}(p,\gamma)^{14}{\rm N}$ reaction proceeds mainly through the $E1$ 
transition from the $1^-$ resonance of the $p+^{13}$C system peaked at about 0.51 MeV 
above the proton threshold of $^{14}$N to $^{14}$N$_{\rm g.s.}$. 
At first glance, it seems straightforward to apply the same folding potential model 
used above for the $^{12}{\rm C}(p,\gamma)$ reaction to study the 
$^{13}{\rm C}(p,\gamma)$ reaction. However, the nonzero spin of the $^{13}$C target 
implies, in general, a spin-spin interaction term in the $p+^{13}$C potential for both 
the scattering and bound states, which has not been taken into account so far in the 
potential model studies of this reaction \cite{Hua10}. 
Concerning the folding model of the \pA potential, the inclusion of a spin-spin 
interaction term complicates significantly the folding procedure that involves the 
spin dependent contribution of the nuclear density. Due to this technical reason, 
the spin-spin interaction term is also widely neglected in the folding model studies 
of elastic \nA scattering as well as \pG reactions. For elastic nucleon scattering, 
the spin-dependent contribution from one or two valence nucleons to the total OP 
is rather small compared to the spin-saturated contribution from the remaining 
nucleons in the target. This is clearly not the case for the  $^{13}{\rm C}(p,\gamma)$ 
reaction, where the $1^-$ resonance is mainly formed by coupling the s$_{1/2}$
scattering state of incident proton to $^{13}$C$_{\rm g.s.}(1/2^-)$ 
\cite{Hua10,Nesa01,King94}. In the ``valence nucleon + core'' approximation, 
this $1^-$ resonance is formed by coupling the s$_{1/2}$ incident proton to 
the p$_{1/2}$ valence neutron in $^{13}$C$_{\rm g.s.}(1/2^-)$, denoted shortly as 
$[\pi_{1/2^+}\otimes\nu_{1/2^-}]_{1^-}$, and $^{14}$N$_{\rm g.s.}(1^+)$ is formed 
by coupling the bound 1p$_{1/2}$ valence proton to the $^{13}$C core. Thus, the 
resonance $\gamma$ transition is the $E1$ transition 
$[\pi_{1/2^+}\otimes\nu_{1/2^-}]_{1^-}\to [\pi_{1/2^-}\otimes\nu_{1/2^-}]_{1^+}$. 
In general, the nonresonant $E1$ transition 
$[\pi_{1/2^+}\otimes\nu_{1/2^-}]_{0^-}\to [\pi_{1/2^-}\otimes\nu_{1/2^-}]_{1^+}$
is also possible, and the wave function of the $1^-$ resonance should be different 
from that of the $0^-$ scattering wave due to the spin-spin interaction. 
Because the spin-spin interaction term is neglected in the present folding model 
approach, the wave functions of both the $1^-$ resonance and $0^-$ scattering state
are generated by the same folded $p+^{13}$C potential, and their difference
is approximately taken into account by using different $N_s$ factors (\ref{eq20}).

\begin{figure}[bht]\vspace*{-0.5cm}
\centerline{\includegraphics[width=0.9\textwidth]{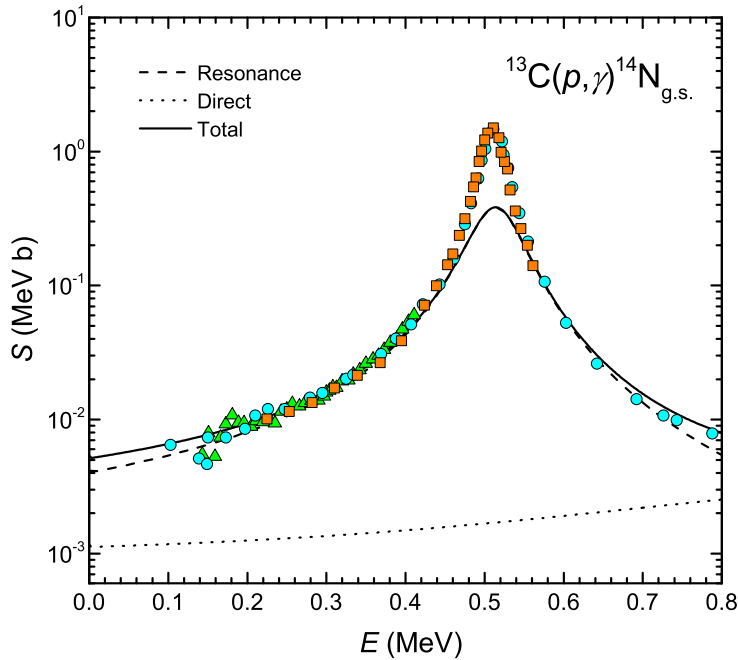}}\vspace*{-0.5cm}
\caption{Astrophysical $S(E)$ factors (\ref{eq1}) of the $^{13}{\rm C}(p,\gamma)^{14}{\rm N}$ 
 reaction given by the folded $p+^{13}$C potential. The best-fit spectroscopic factor 
 obtained for the $[\pi_{1/2^+}\otimes\nu_{1/2^-}]_{1^-}$ resonance radiative capture   
 is $S_{\rm F}\approx 0.23$, with that for the direct $[\pi_{1/2^+}\otimes\nu_{1/2^-}]_{0^-}$
 nonresonant radiative capture $S_{\rm F}=0.69$ taken from the shell model results 
 \cite{Coh67}. The experimental data shown as circles, triangles, and squares
 were taken from Refs.~\cite{King94}, \cite{Heb60}, and \cite{Gen10}, respectively.} 
 \label{f5}
\end{figure}
For the bound 1p$_{1/2}$ proton state of $^{14}$N, using the same spin-orbit potential 
as given above in Sec.~\ref{sec4}, the folded $p+^{13}$C potential calculated at the 
experimental binding energy $E=-E_b$ ($E_b\approx 7.55$ MeV) needs to be scaled 
by $N_{b}\approx 1.105$ to reproduce the eigenvalue $E=-E_b$ in the solution of 
Eq.~(\ref{eq2}). This factor is about 10\% larger than unity and this might be caused
by the lack of the spin-spin interaction term in Eq.~(\ref{eq2}). The $1^-$ resonance 
peak observed at $E\approx 0.51$ MeV \cite{King94,Heb60,Gen10} is reproduced by the 
folded $p+^{13}$C potential (\ref{eq20}) as the $[\pi_{1/2^+}\otimes\nu_{1/2^-}]_{1^-}$ 
peak using $N_{s}\approx 1.20$, with the statistic $\chi^2\approx 0.02$ for one degree of freedom. The best-fit 
spectroscopic factor obtained for  the radiative capture $^{13}{\rm C}(p,\gamma)$ 
to this $1^-$ resonance is $S_{\rm F}\approx 0.23$ (see Fig.~\ref{f5}). 

Because the resonance peak at $E\approx 0.51$ MeV contains no contribution from
the $0^-$ scattering wave (only a very weak $0^-$ peak was observed so far in the 
$^{13}{\rm C}(p,\gamma)^{14}{\rm N}$ cross section at $E\approx 1.2$ MeV \cite{Cha15}),
we have used $N_{s}\approx 1.08$ for the $p+^{13}$C folded potential to generate 
the nonresonant $0^-$ scattering states based on a good fit of the total astrophysical 
$S$ factor to the data (see Fig.~\ref{f5}). The spectroscopic factor $S_{\text{F}}=0.69$ 
taken from the shell model results \cite{Coh67} was used to obtain the nonresonant direct
capture cross section by the $0^-$ scattering waves. As noted above, the absence of the 
spin-spin interaction term results on different renormalization factors of the folded 
$p+^{13}$C potential for the nonresonant $0^-$ scattering ($N_{s}\approx 1.08$) and 
$1^-$ resonance  ($N_{s}\approx 1.20$). The final state ANC given by our folding model 
calculation is $A_{\rm F}^2\approx 15.24$~fm$^{-1}$, which is about 60\% larger 
than that deduced in Ref.~\cite{Hua10} but quite close to the values 
$A_{\rm F}^2\approx 18.6$ and 17.8~fm$^{-1}$ deduced from the DWBA analysis 
of the nucleon transfer reaction $^{13}$C($^{14}$N,$^{13}$C)$^{14}$N measured at 
the incident energy 162 MeV \cite{Tra98} and $^{13}$C($^3{\rm He},d)^{14}$N measured 
at 26.3 MeV \cite{Bem00}, respectively. Based on these nucleon transfer data, an averaged 
value $A_{\rm F}^2\approx 18.2$ fm$^{-1}$ was adopted in the $R$-matrix calculation 
of the $^{13}{\rm C}(p,\gamma)^{14}{\rm N}$ reaction by Mukhamedzhanov {\it et al.} 
\cite{Muk03}. Similar DWBA analysis of the $^{13}$C($^3{\rm He},d)^{14}$N reaction 
measured at incident energies of 22.3, 32.5 and 34.5 MeV by Artemov \textit{et al.} 
obtained $A_{\rm F}^2\approx 16.5_{-2.3}^{+2.6}$ fm$^{-1}$ \cite{Art08}. It is 
noteworthy that Chakraborty \textit{et al.} have treated the final state ANC as 
free parameter in their $R$-matrix analysis of the $^{13}{\rm C}(p,\gamma)^{14}{\rm N}$ 
reaction and found $A_{\rm F}^2\approx 15.89$ fm$^{-1}$ \cite{Cha15} which agrees 
nicely with our result.

\begin{figure}[bht]\vspace*{-0.5cm}
	\centerline{\includegraphics[width=\textwidth]{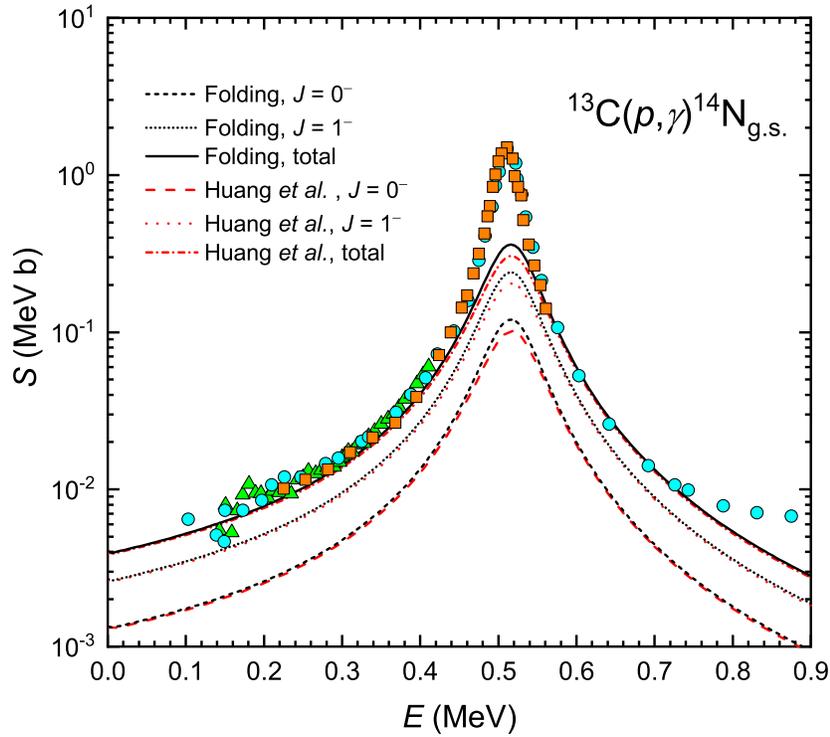}}\vspace*{-0.5cm}
	\caption{Astrophysical $S(E)$ factors (\ref{eq1}) of the $^{13}{\rm C}(p,\gamma)^{14}{\rm N}$ 
reaction given by the folded $p+^{13}$C potential renormalized by $N_{s}\approx 1.20$ 
for both $[\pi_{1/2^+}\otimes\nu_{1/2^-}]_{1^-}$ and $[\pi_{1/2^+}\otimes\nu_{1/2^-}]_{0^-}$
scattering states, with the best-fit spectroscopic factor $S_{\text{F}}\approx 0.15$. 
The similar results obtained with the WS potential parametrized by Huang {\it et al.} 
\cite{Hua10} are shown for the comparison. The experimental data are the same as 
those shown in Fig.~\ref{f5}.} \label{f6}
\end{figure}
About the same $S(E)$ factor of the $^{13}{\rm C}(p,\gamma)^{14}{\rm N}$ reaction
was obtained by Huang {\it et al.} \cite{Hua10} using the phenomenological WS 
potential, but their best-fit spectroscopic factor ($S_{\rm F}\approx 0.15$) is 
nearly 50\% smaller than that given by the present folding model study. The reason 
for such a difference is that the authors of Ref.~\cite{Hua10} have assumed the 
resonance peak at $E\approx 0.51$ MeV as a mixture of both the $1^-$ and 
$0^-$ scattering states and adjusted WS parameters to generate the resonance 
wave function by coupling the s$_{1/2}$ incident proton to the p$_{1/2}$ 
valence neutron in $^{13}$C target to the total spin $1^-$ and $0^-$. 
Without the spin-spin interaction term, the obtained radial wave functions 
of these two resonance scattering waves are exactly the same and their total 
strength needs to be scaled only by $S_{\rm F}\approx 0.15$ for the best fit 
to the measured data. It is well established that the first $0^-$ excitation 
of $^{14}$N lies at 1.2 MeV above the proton threshold \cite{A-Love91}, and it 
was observed only as a weak peak in the $^{13}{\rm C}(p,\gamma)$ cross section 
at $E\approx 1.2$ MeV \cite{Cha15}. Therefore, we deem the potential model
description of the resonance peak at $E\approx 0.51$ MeV given in Ref.~\cite{Hua10}
as unrealistic. The nonresonant contribution of the direct $^{13}{\rm C}(p,\gamma)$
capture was included in this study by using another WS potential, and 
$A_{\rm F}^2\approx 9.3$ fm$^{-1}$ was obtained \cite{Hua10} which is significantly
smaller than the $A_{\rm F}^2$ values quoted above. To illustrate this effect, we 
have used the renormalization factor $N_{s}=1.20$ for the folded $p+^{13}$C potential 
to describe the resonance peak at $E\approx 0.51$ MeV as a mixture of both 
the $1^-$ and $0^-$ resonance scattering waves, and found the best-fit  
$S_{\text{F}} \approx 0.15$ which is the same as that 
obtained in Ref.~\cite{Hua10} (see Fig.~\ref{f6})

\begin{figure}[bht]\vspace*{-1.5cm}
\centerline{\includegraphics[width=\textwidth]{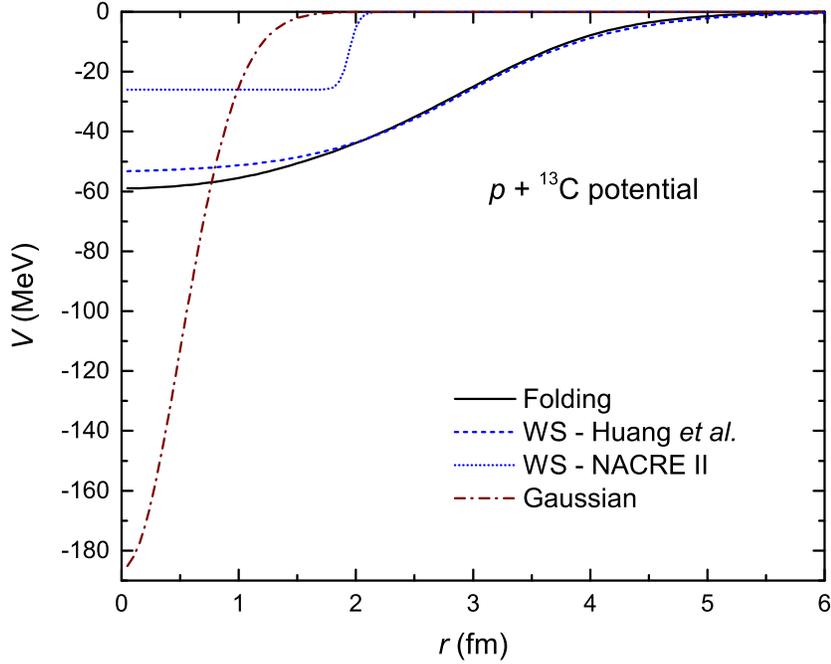}}\vspace*{-0.5cm}
\caption{Different $p+^{13}$C potentials used to generate the resonance  
$[\pi_{1/2^+}\otimes\nu_{1/2^-}]_{1^-}$ scattering state: the folded potential 
renormalized by $N_{s}=1.20$, WS potential parametrized by Huang {\it et al.} 
\cite{Hua10} and that taken from NACRE II compilation \cite{NACRE2}, and the deep 
Gaussian potential taken from Ref.~\cite{Dub12}.} \label{f7}
\end{figure}
To fill in the sharp resonance peak observed at $E\approx 0.51$ MeV remains a 
problem for the present folding model study. So far, this sharp peak  
could be filled only by using a shallow WS potential with an extremely small 
diffuseness, $V_{\rm N}= 25.82$ MeV, $R_{\rm N}=1.944$ fm, and $a_{\rm N}=0.04$ 
fm \cite{NACRE2}, or using a very deep and narrow Gaussian potential 
$V(r)=V_0\exp(-\beta r^2)$ with $V_0=-186.07$~MeV and $\beta=2$~fm$^{-2}$ \cite{Dub12}. 
Fig.~\ref{f7} shows different $p+^{13}$C potentials used to obtain the resonance
$[\pi_{1/2^+}\otimes\nu_{1/2^-}]_{1^-}$ scattering state (note that the spin-orbit
contribution is zero for the s wave), and one can see that the best-fit folded 
potential and WS potential used by Huang {\it et al.} \cite{Hua10} are close 
to the mean-field based \nA OP at low energies, with the depth around 
$50\sim 60$ MeV \cite{Kon03}. The WS potential taken from NACRE II compilation 
\cite{NACRE2} and the deep Gaussian potential \cite{Dub12} are unrealistic 
and become zero already at $r\approx 2$ fm. Such a short distance of the
$p+^{13}$C interaction is well in the interior of the $^{13}$C nucleus whose 
charge radius is $R_{\rm C}\approx 2.5$ fm \cite{Hei70,Sch82}, and the 
nuclear interaction there should not be neglected at all.

\begin{figure}[th]\vspace*{-1.0cm}
\centerline{\includegraphics[width=\textwidth]{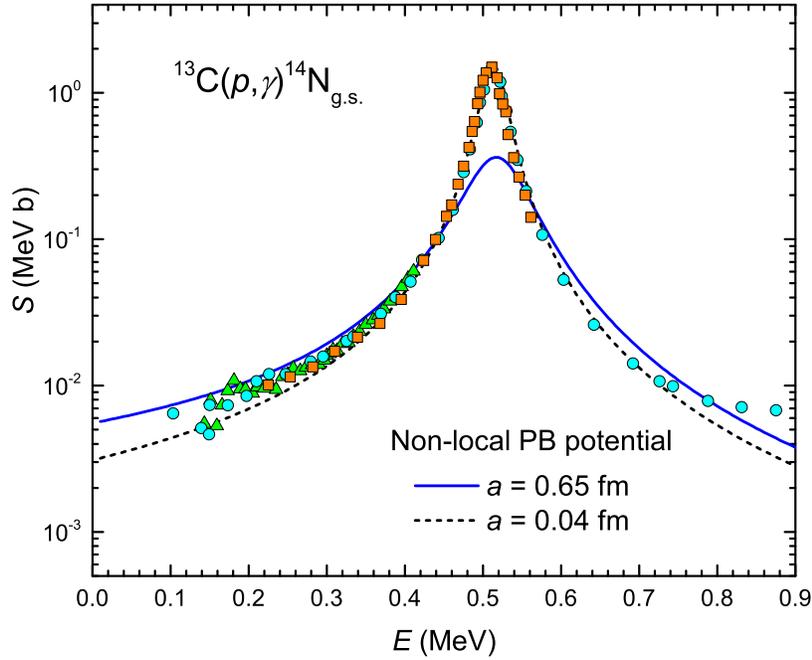}}\vspace*{-0.5cm}
\caption{Astrophysical $S(E)$ factors (\ref{eq1}) of the $^{13}{\rm C}(p,\gamma)^{14}$N
reaction obtained using the nonlocal PB-type potential for the resonance scattering. 
The solid and dotted lines are the results obtained with the diffuseness $a=0.65$ fm 
and $a=0.04$ fm, respectively. The experimental data are the same as those shown in 
Fig.~\ref{f5}.}  \label{f8}
\end{figure}

Recently, Tian \textit{et al.} \cite{Tian18} have used the nonlocal potential of 
the Perey-Buck (PB) form to determine the astrophysical $S$ factors for some 
radiative capture reactions using the potential parameters $R_{\rm N}=1.25A^{1/3}$ 
fm and $a=0.65$ fm. To probe possible effect caused by a very small diffuseness $a$,
we have performed similar potential model calculation of the $S$ factor using 
the nonlocal PB-type potential. Using the same PB parameters as those adopted
in Ref.~\cite{Tian18}, we varied the potential depth slightly to reproduce 
the $p+^{13}$C resonance peak at $E\approx 0.51$ MeV and the experimental proton 
separation energy $E_b$ of $^{14}$N. As a result, we found $V_N = 61.89$ MeV 
for the proton bound state in $^{14}$N and $V_N = 69.50$ MeV for the $p+^{13}$C 
resonance at 0.51 MeV, with the best-fit $S_{\rm F}\approx 0.22$. To explore 
the effect of a small diffuseness, the parameters of PB potential similar to those 
taken from NACRE II compilation \cite{NACRE2} were also used, namely, $R = 1.944$ fm 
and $a = 0.04$ fm. Keeping the range of nonlocality at $\beta_N =0.85$ fm, we 
obtained $V_N = 124.75$ MeV for the bound state and $V_N = 27.91$ MeV for the 
resonance scattering, and the best-fit astrophysical factor $S_{\rm F} \approx 0.40$. 
The results given by the nonlocal $p+^{13}$C potential of the FB form are shown 
in Fig.~\ref{f8}, and one can see that the sharp resonance peak cannot be reproduced 
by the nonlocal PB potential with $a = 0.65$ fm. The sharp resonance peak at 
$E\approx 0.51$ MeV can be well reproduced only by a nonlocal PB potential with 
the shallow depth $V_{\rm N}= 27.91$ MeV and extremely small diffuseness $a = 0.04$ 
fm, like those of the shallow WS potential taken from NACRE II compilation 
\cite{NACRE2}. Given the spin-spin interaction {\it not} included in the potential 
model calculations discussed above, it is not excluded that the inclusion of this 
term into the $p+^{13}$C potential might help to reproduce the sharp resonance 
peak observed at $E\approx 0.51$ MeV.

\begin{figure}\vspace*{-1.5cm}
 \centerline{\includegraphics[width=0.8\textwidth]{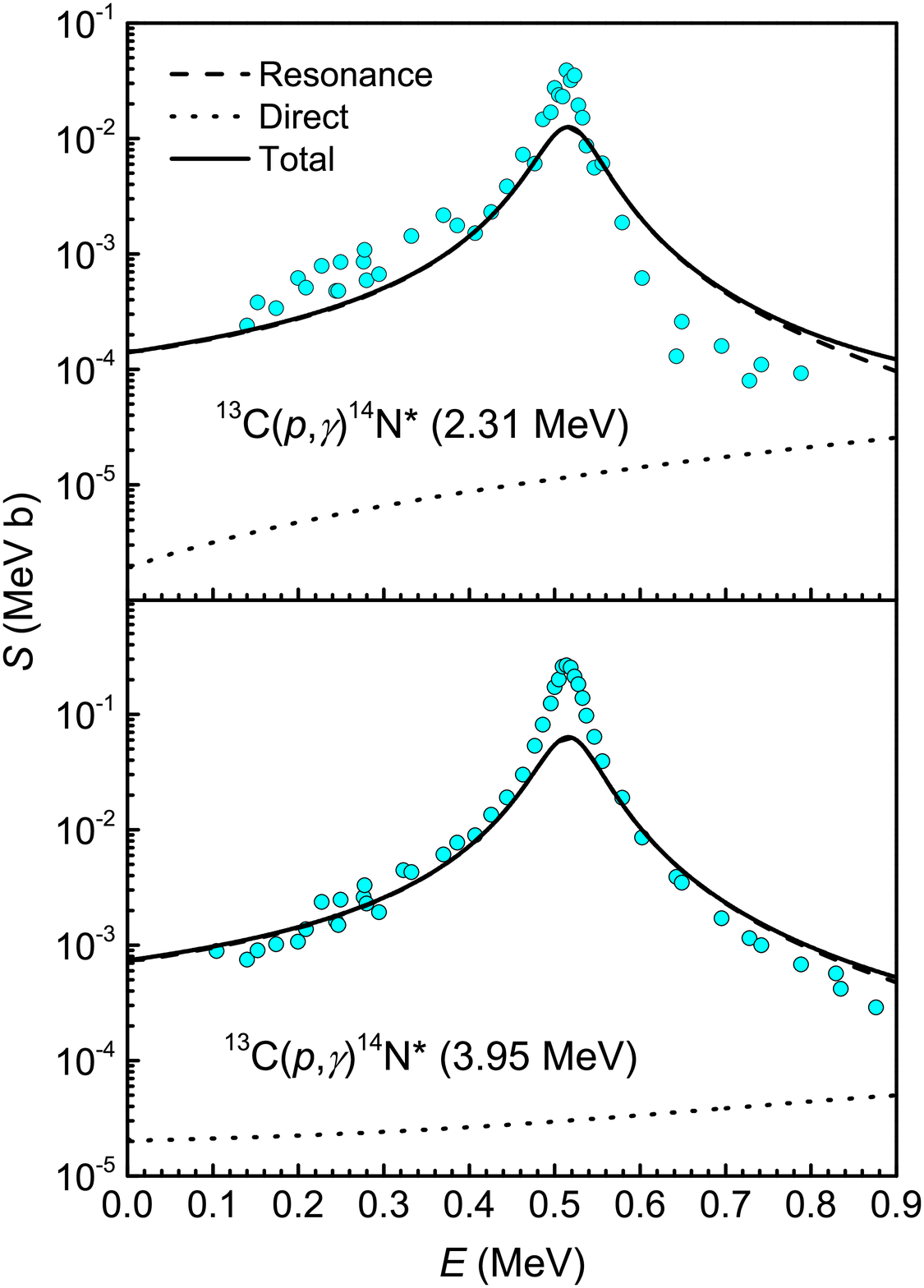}}\vspace*{-0.5cm}
\caption{ Astrophysical $S(E)$ factors (\ref{eq1}) of the 
$^{13}{\rm C}(p,\gamma)^{14}{\rm N^*}$ reaction given by the folded $p+^{13}$C 
potential, with $^{14}\rm N^*$ being in the $0^+$ excited state at 2.31 MeV 
(upper panel) and $1^+$ excited state at 3.95 MeV (lower panel). The data were 
taken from Ref.~\cite{King94}.}\label{f9}
\end{figure}
We show finally that our folding model approach can also be used to describe 
the \pG cross sections for the excited states of $^{14}$N \cite{King94}. We focus 
here on the electric dipole $\gamma$ transitions feeding the $0^+$ (2.31 MeV) and 
$1^+$ (3.95 MeV) excited states of $^{14}$N. In the simple ``valence nucleons + core'' 
scenario, the $0^+$ and $1^+$ excited states of $^{14}$N are formed by coupling the 
bound p$_{1/2}$ proton to the p$_{1/2}$ neutron state,  
$[\pi_{1/2^-}\otimes\nu_{1/2^-}]_{0^+,1^+}$ \cite{Muk03,Art08}. 
Because these $E1$ transitions proceed mainly from the $1^-$ resonance 
at $E\approx 0.51$ MeV, the factor $N_{s}\approx 1.20$ was also used for 
the folded $p+^{13}$C potential to generate the resonance  
$[\pi_{1/2^+}\otimes\nu_{1/2^-}]_{1^-}$ scattering wave. For the bound p$_{1/2}$
proton of $^{14}$N being in the $0^+$ and $1^+$ excited states, the folded $p+^{13}$C 
potential was scaled by $N_{b}\approx 1.022$ and 0.959 to reproduce the proton 
separation energies $E_b\approx 5.24$ and 3.6 MeV, respectively. The obtained 
astrophysical $S(E)$ factor for the $^{13}{\rm C}(p,\gamma)^{14}$N$^*_{0^+,1^+}$ 
reactions agree reasonably with the data as shown in Fig.~\ref{f9}, with the 
best-fit $S_{\rm F}\approx 0.032$ and 0.17 for the $0^+$ and $1^+$ excited states,
respectively.  

Besides the resonance radiative capture, the direct (nonresonant) capture is possible 
through the $E1$ transitions from the nonresonant $[\pi_{3/2^+}\otimes\nu_{1/2^-}]_{1^-}$ 
and $[\pi_{1/2^+}\otimes\nu_{1/2^-}]_{0^-}$ scattering states to the $0^+$ and $1^+$ 
excited states of $^{14}$N, respectively.
To determine the nonresonant $1^-$ scattering wave, the folded $p+^{13}$C potential 
was renormalized by the same factor $N_{s}\approx 1.20$ as that used to determine 
the $1^-$ resonance scattering state. For the nonresonant $0^-$ scattering wave,
we have used $N_{s}\approx 1.08$ as done above for the direct (nonresonant) capture 
through the $0^-$ scattering state to $^{14}$N$_{\rm g.s.}$. Using 
$S_{\text{F}} \approx 0.87$ taken from the shell model results \cite{Coh67}, 
the final state ANC of the direct capture $^{13}{\rm C}(p,\gamma)^{14}$N$^*_{0^+}$ 
reaction was obtained as $A_{\rm F}^2\approx 10.49$~fm$^{-1}$ which is close to 
those used in the $R$-matrix calculations of this reaction, 
$A_{\rm F}^2\approx 8.90$~fm$^{-1}$ \cite{Muk03}, 
$A_{\rm F}^2\approx 12.2^{+1.7}_{-1.9}$~fm$^{-1}$ \cite{Art08}, and 
$A_{\rm F}^2\approx 8.84$~fm$^{-1}$ \cite{Cha15}.
Using $S_{\text{F}}\approx 0.035$ taken from the shell model results \cite{Coh67}, 
we obtained the final state ANC of the (nonresonant) direct capture 
$^{13}{\rm C}(p,\gamma)^{14}$N$^*_{1^+}$ reaction as 
$A_{\rm F}^2\approx 0.252$~fm$^{-1}$ which is smaller than 
those used in the $R$-matrix calculations of this reaction, 
$A_{\rm F}^2\approx 2.71$~fm$^{-1}$ \cite{Muk03} and 
$A_{\rm F}^2\approx 2.14^{+0.32}_{-0.30}$~fm$^{-1}$ \cite{Art08}. 
Such a discrepancy might be due to a rather small $S_{\rm F}$ value of the 
bound 1p$_{1/2}$ proton state given by the shell model calculation \cite{Coh67}. 
We note here a very small value of $S_{\rm F} = 0.0015$ used in the NACRE II 
evaluation of this reaction \cite{NACRE2} which is likely associated with 
the choice of proton binding potential. 

We conclude from the results shown in Figs.~\ref{f3}, \ref{f5}, and \ref{f9} that 
the contribution from the nonresonant direct capture to the total \CpG cross 
section is negligible compared to the contribution from the resonance proton
capture in the considered energy range.

\section{Summary}
The local proton OP given by the mean-field based folding model 
\cite{Kho02,Loan15} has been used consistently in the OM analysis of elastic 
\pC23 scattering at low energies and in the potential model study of the 
\CpG reactions. The elastic $p+^{12,13}{\rm C}$ scattering data at energies 
around the Coulomb barrier are well described by the folded proton OP 
renormalized by $N_s\approx 1.2\sim 1.3$, quite close to that required for 
the good folding model description of the resonance \CpG capture. 
This same folding model was also used as the binding potential 
of the proton bound state in daughter nuclei of the \CpG reactions, 
with the best-fit renormalization coefficient $N_b$ of the folded \pC23 
potential close to unity. 

In the ``valence proton + core'' approximation for the $p+^{12}$C system, 
the reduced $E1$ transition probability (\ref{eq22}) calculated using the 
proton scattering- and bound-state wave functions given by the folded 
$p+^{12}$C potential agrees well with $B_{\rm exp}(E1;1/2^+\to {\rm g.s.})$ 
adopted for the $1/2^+$ excited state of $^{13}$N. A good folding model 
description of the astrophysical $S(E)$ factor of both the resonance and
direct (nonresonant) $^{12}$C$(p,\gamma)$ capture reaction has been obtained,
and the best-fit spectroscopic factor $S_{\rm F}$ and ANC are comparable
with those deduced earlier using the phenomenological WS potential. 

For the $^{13}{\rm C}(p,\gamma)$ reaction, the $E1$ transitions 
from the $1^-$ resonance at $E\approx 0.51$ MeV to the ground state as well 
as the $0^+$ and $1^+$ excited states of $^{14}$N have been consistently 
studied, and the best-fit spectroscopic factors and ANC's obtained for the 
$^{13}{\rm C}(p,\gamma)$ capture reaction are compared with those deduced
from the analysis of proton transfer reactions and/or used in the
$R$-matrix calculation of this reaction. 

Although the calculated astrophysical $S(E)$ factor agrees reasonably with 
data measured for the $^{13}{\rm C}(p,\gamma)$ capture reaction, the narrow 
resonance peak at $E\approx 0.51$ MeV could not be reproduced by our folding model 
approach. The sharp peak could be reproduced only by a very short-range shallow 
WS potential with an extremely small diffuseness \cite{NACRE2} that is unrealistic 
from the mean-field point of view. The same conclusion was drawn from the results 
obtained with the nonlocal $p+^{13}$C potential of the Perey-Buck form.  

The present work shows the validity of the mean-field based folding model of the 
\pA potential that the can be consistently used to describe the elastic proton 
scattering at low energies and \pG reactions of interest for nuclear astrophysics. 
Given the nonlocal version of the microscopic folding model developed recently 
to describe elastic nucleon scattering \cite{Loan20}, its further use to describe 
\pG reactions is planned for a future study. The inclusion of the spin-spin 
interaction into the potential model description of the \pG capture reaction 
by a nonzero-spin target would also be an important development of the model.    

\section*{Acknowledgements}
We thank Pierre Descouvemont for his communication on the numerical calculation 
of the \pG cross section, Bui Minh Loc and Doan Thi Loan for their assistance 
in the folding model calculation of the \pC23 potential. The DOLFIN code used 
to generate the nuclear densities in the IPM has been provided to one of us 
(D.T.K.) by the late Ray Satchler. The present research was supported, in part, 
by Vietnam Atomic Energy Institute (VINATOM) under the grant 
{\DJ}TCB.01/19/VKHKTHN. 

\bibliographystyle{model1-num-names}
\bibliography{cas-refs}
\end{document}